\documentclass[%
reprint,
superscriptaddress,
twocolumn,
amsmath,
amssymb,
aps,
prb
]{revtex4-1}

\usepackage{graphicx}  
\usepackage{dcolumn}  
\usepackage{bm}            
\usepackage{bbold}
\usepackage{amsmath}
\usepackage{xfrac}
\usepackage{xcolor}
\usepackage[colorlinks=true, citecolor=red]{hyperref}   
\usepackage{hhline}
\usepackage{diagbox}
\usepackage{siunitx}
\usepackage{cancel}
\usepackage{enumitem}
\usepackage[normalem]{ulem}

\newcommand{\kOhm}{\textrm{\SI{}{\kohm}}}
\newcommand{\kohm}{k$\Omega$~}
\newcommand{\Ohm}{\textrm{\SI{}{\ohm}}}

\usepackage{siunitx}

\usepackage{tabularx}
\usepackage{array}
\newcommand\redsout{\bgroup\markoverwith{\textcolor{red}{\rule[0.5ex]{2pt}{1.0pt}}}\ULon}

\newcommand{\corev}[1]{{\color{black}#1}}

\begin{document}
	
	\title{Thermal crossover from a Chern insulator to a fractional Chern insulator in pentalayer graphene}
	
	\author{Sankar Das Sarma}
	\affiliation{Condensed Matter Theory Center and Joint Quantum Institute, Department of Physics, 
		University of Maryland, College Park, Maryland 20742, USA}

	\author{Ming Xie}
	\affiliation{Condensed Matter Theory Center and Joint Quantum Institute, Department of Physics, 
		University of Maryland, College Park, Maryland 20742, USA}

	\date{\today}
	
	\begin{abstract}
		
		By theoretically analyzing the recent temperature dependent transport data 
		[Lu \textit{et al}., arXiv:2408.10203] in pentalayer graphene, 
		we establish that the experimentally observed transition 
		from low-temperature quantum anomalous Hall effect (QAHE)
		to higher-temperature fractional quantum anomalous Hall effect (FQAHE)
		is a crossover phenomenon arising from
		the competition between interaction and disorder energy scales, 
		with the likely zero temperature ground state being 
		either a localized insulator or a Chern insulator with a quantized anomalous Hall effect.  
		In particular, the intriguing suppression of FQAHE in favor of QAHE with decreasing temperature is explained 
		as arising from the low-temperature localization of the carriers where disorder overcomes the interaction effects.
		We provide a detailed analysis of the data in support of the crossover scenario.

	\end{abstract}
	
	\maketitle

In a recent follow-up experiment \cite{JuEQAH2024} 
to their breakthrough discovery of quantum anomalous Hall (QAHE) 
and fractional quantum anomalous Hall (FQAHE) effects in pentalayer graphene (PLG) \cite{PentaGraphene2023},
Ju and collaborators at MIT made a startling discovery that, 
with the lowering of temperature ($T$) from above $300$ mK to below $100$ mK, 
the original PLG FQAHE reported in Ref.~\onlinecite{PentaGraphene2023}
is suppressed in favor of a QAHE phase, which spans across a large range of the filling factor ($\nu$) 
in some range of the applied displacement field ($D$).  
In addition, as $T$ decreases, the strongly insulating (SI) phase, 
also observed in some low-$\nu$ corners of the $D-\nu$ space in the original paper\cite{PentaGraphene2023}, 
expands along with the QAHE phase.  
The purpose of the current paper is to analyze the new low-$T$ data in Ref.~\onlinecite{JuEQAH2024} and 
suggest possible physical explanations for the puzzling phenomena reported therein.  
Our analyses of the experiment use the actual data underlying Ref.~\onlinecite{JuEQAH2024}, 
which were supplied to us  by the authors of Ref.~\onlinecite{JuEQAH2024}. 

\begin{figure}[b!]
 	\includegraphics[width=0.38\textwidth]{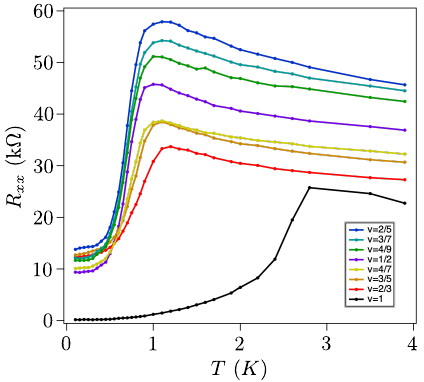}
 	\caption{\label{Rxxdata}Temperature dependence of the longitudinal resistance $R_{xx}$ at integer and fractional
 		filling factors. The data are taken from device 1 reported in Ref.~\onlinecite{PentaGraphene2023} (courtesy of Ju et al.).
 	}
 \end{figure}

We earlier analyzed \cite{XiePenta2024} the higher-$T$ results reported in Ref.~\onlinecite{PentaGraphene2023}, 
which can be summarized as follows.  
We found a seemingly puzzling constant activation gap ($\Delta_{\nu}\negmedspace\sim\negmedspace5$ K) 
for all the reported FQAHE fractions, $\nu=2/5,3/7,4/9,4/7,3/5, 2/3$ 
along with a QAHE gap $\sim10$ K for $\nu=1$.  
In addition, the FQAHE for all $\nu$ generically manifest a mysterious residual 
longitudinal resistance $R_0>10$ \kOhm~at low temperatures, 
which is unheard of in the quantum Hall literature, where the quantization of $R_{xy}$ is always associated with the vanishing of $R_{xx}$.  
By contrast, the QAHE at $\nu=1$ shows a much smaller value of $R_0\sim 0.1-0.5\,\kOhm$.
There was an additional mystery in the original experiment \cite{PentaGraphene2023}, as shown in Fig.~\ref{Rxxdata}.  
The $T$-dependence of $R_{xx}$ for all fractions is essentially identical for $\nu \in [2/5,2/3]$
including at $\nu=1/2$, which is extremely puzzling,
while $R_{xx}(T)$  is qualitatively different  at $\nu=1$.  
We note that Ref.~\onlinecite{JuEQAH2024} basically manifests an SI phase for $\nu<2/5$ 
(e.g., $\nu=1/3$ does not manifest any FQAHE as it falls within the strongly insulating regime in both Refs.~\onlinecite{JuEQAH2024,PentaGraphene2023}), 
so only the regime $\nu>2/5$ is relevant for our consideration.

The results in Fig.~\ref{Rxxdata} clearly bring out the two essential features of the data in Ref.~\onlinecite{PentaGraphene2023}, namely, 
the constancy of the activation energy for all fractional $\nu$ values
 (as reflected in the almost parallel rise in the activated $R_{xx}$ for $0.3$ K$<T<1$ K) including $\nu=1/2$, 
 and the existence of a large $R_0$ for all the fractions as reflected in $R_{xx} (T<0.3K)$ approaching a constant value $R_0 \sim 10-15\,\kOhm$.
 By contrast, $\nu=1$ QAHE produces a very small $R_0$ in Fig.~\ref{Rxxdata}.  
 Another surprising feature of Fig.~\ref{Rxxdata}, which was not reported in Ref.~\onlinecite{XiePenta2024} 
 (and was entirely missed in Ref.~\onlinecite{PentaGraphene2023}), 
 is that the data points to an activated temperature dependence of $R_{xx}$ at $\nu=1/2$, 
 with a gap $\Delta_{1/2} \sim 5$ K (and $R_0 \sim 12$ k$\Omega$), implying an apparent FQAHE at $\nu=1/2$.
The corresponding Hall resistance $R_{xy}$, however, reflects no plateau at $\nu=1/2$ in Ref.~\onlinecite{PentaGraphene2023}.
 This apparent conflict remained a mystery for the data of 
 Ref.~\onlinecite{PentaGraphene2023} and was therefore not mentioned in Ref.~\onlinecite{XiePenta2024}.

The new experiment \cite{JuEQAH2024} now sheds some light on this mystery
by extending the temperature range down to $T\sim 40$ mK
(the reliable $T$ values in Ref.~\onlinecite{PentaGraphene2023} were above $200\text{--}300$ mK). 
The new discovery of Ref.~\onlinecite{JuEQAH2024} is stunning: 
With decreasing $T$, the QAHE phase takes over three nearly connected $\nu\text{--}D$ regions, 
extending from $\nu=1/2$ to above $\nu=1$,
and FQAHE states in these regions are suppressed with the
$R_{xy}$ now systematically becoming $h/e^2$ integer quantization!  
The corresponding $R_{xx }(T)$ shows a crossover behavior (in contrast to a clear activated behavior) 
until $T$ is low enough to reach the QAHE state.
In addition to these extended regions of QAHE, 
the SI regime also expands appreciably covering much of the $\nu<1/2$ region 
where both $R_{xx}$ and $R_{xy}$ become very large at low $T$.  
The experimental results of Ref.~\onlinecite{JuEQAH2024} are consistent with the system (at some $D$ values) 
showing an extended quantum anomalous Hall effect (EQAHE) 
over a large range of $\nu$ values 
 with $R_{xy}$ being quantized at $h/e^2$ as in the regular QAHE.
The EQAHE appears only as $T$ decreases,  since for $T \gtrsim 0.3K$ the QAHE manifests only at $\nu\sim 1$
and the FQAHE is observed for the usual Jain fractions \cite{Jain1989} of $\nu=2/3, 3/5$, etc, as in Ref.~\onlinecite{PentaGraphene2023}.  
In fact, the results of Ref.~\onlinecite{JuEQAH2024} seem to indicate that the $T=0$ phase is likely to be EQAHE at all $\nu$ values except  for $\nu<2/5$ which is SI.  

\begin{figure}[t!]
	\includegraphics[width=0.4\textwidth]{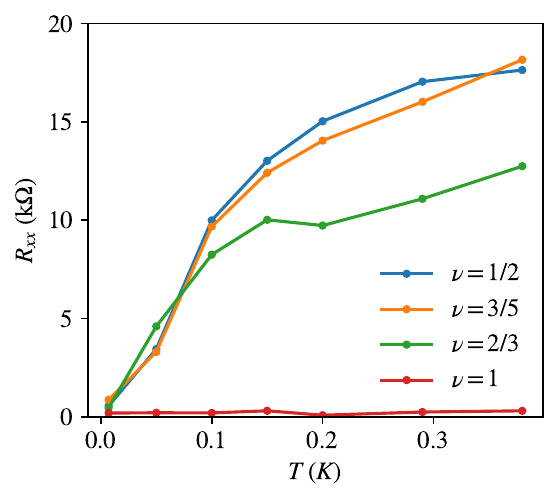}
	\caption{\label{LowTRxx}Temperature dependence of the longitudinal resistance $R_{xx}$ for the same Device 1 as in Fig.~\ref{Rxxdata} but at lowered temperatures
		extracted from Ref.~\onlinecite{JuEQAH2024} (courtesy of Ju et al.).
	}
\end{figure}

\corev{
	In this work, we combine previous findings \cite{PentaGraphene2023,XiePenta2024} with the new experimental discovery \cite{JuEQAH2024}
	and show that they converge on a coherent physical picture of thermal crossover 
	from the high-$T$ FQAHE phase to the low-$T$ EQAHE phase as temperature decreases.
	We argue that the crossover originates from the competition between electron-electron interactions, 
	which favor the FQAHE phase, 
	and disorder, which induces the EQAHE state.
	This scenario provides a unified explanation that resolves the mysteries in both the high-$T$ and the low-$T$ experiments.
	In fact, as is already evidenced in Fig.~\ref{Rxxdata},
	the large and universal background	resistance $R_0\sim 10$ \kohm 
	at fractional fillings is unlikely to be a coincidence but may indicate that the FQAHE states 
	occur on top of a universal background state with large resistance ($\sim R_0$) likely caused by disorder.
	Lowering $T$  leads to the true ground state, i.e., the EQAHE state,  
	with small residual $R_{xx}$ and quantized $R_{xy} \sim h/e^2$ due to reflectionless edge conduction.
}

\begin{figure*}[t!]
	\includegraphics[width=0.9\textwidth]{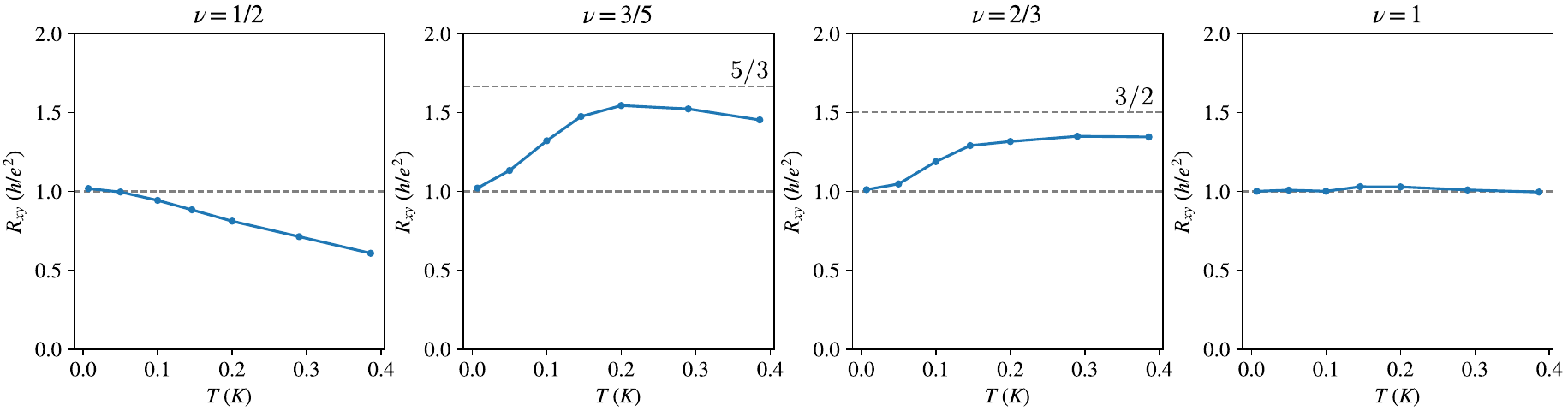}
	\caption{\label{LowTRxy}Temperature dependence of the Hall resistance $R_{xy}$ for Device 1 
		extracted from Ref.~\onlinecite{JuEQAH2024}. (Courtesy of L. Ju, et al.)
		The dashed lines mark the expected integer and fractional quantized values of $R_{xy}$.
	}
\end{figure*}

We begin by showing our analyses of the data of Ref.~\onlinecite{JuEQAH2024}, 
and establishing that the crossover phenomenology described above is consistent with the data.  
In Fig.~\ref{LowTRxx}, we show our extracted $R_{xx} (T)$ from Ref.~\onlinecite{JuEQAH2024} 
for $\nu=1/2, 3/5, 2/3,1$ in the temperature range $T\in (0.01 \textrm{ K}, 0.3\textrm{ K})$.
It is clear that $R_{xx}$ is showing a crossover from one phase ($T<0.1$ K) to a different phase ($T>0.3$ K) with no clear activation behavior.  
It is also manifestly obvious that the low-$T$ `phase'  (up to $T\sim 0.1$ K) is the same for all three fractions 
whereas the high-$T$ ($T>0.3$ K) phase is different for different $\nu$s as expected for FQAHE \cite{footnote}.  
We attribute the low-$T$ phase to the EQAHE phase, which eventually succumbs to the FQAHE phase around $T \sim 0.3$ K.  
We mention that the corresponding $R_{xx} (T)$ for the QAHE  around $\nu=1$
manifests an almost $T$-independent $R_{xx}\sim 0$ throughout the $T\in (0.01 \textrm{ K}, 0.3\textrm{ K})$ range,
 consistent with its activation gap being larger than 10K as seen in Ref.~\onlinecite{PentaGraphene2023}. 
Note that the physics depends on the value of $D$, and Fig.~\ref{LowTRxx} shows results for a specific $D$ 
since this is what is provided in Ref.~\onlinecite{JuEQAH2024}.  The low-$T$ EQAHE 
to the high-$T$ FQAHE crossover depends on the applied displacement field.

\begin{figure}[b!]
	\includegraphics[width=0.35\textwidth]{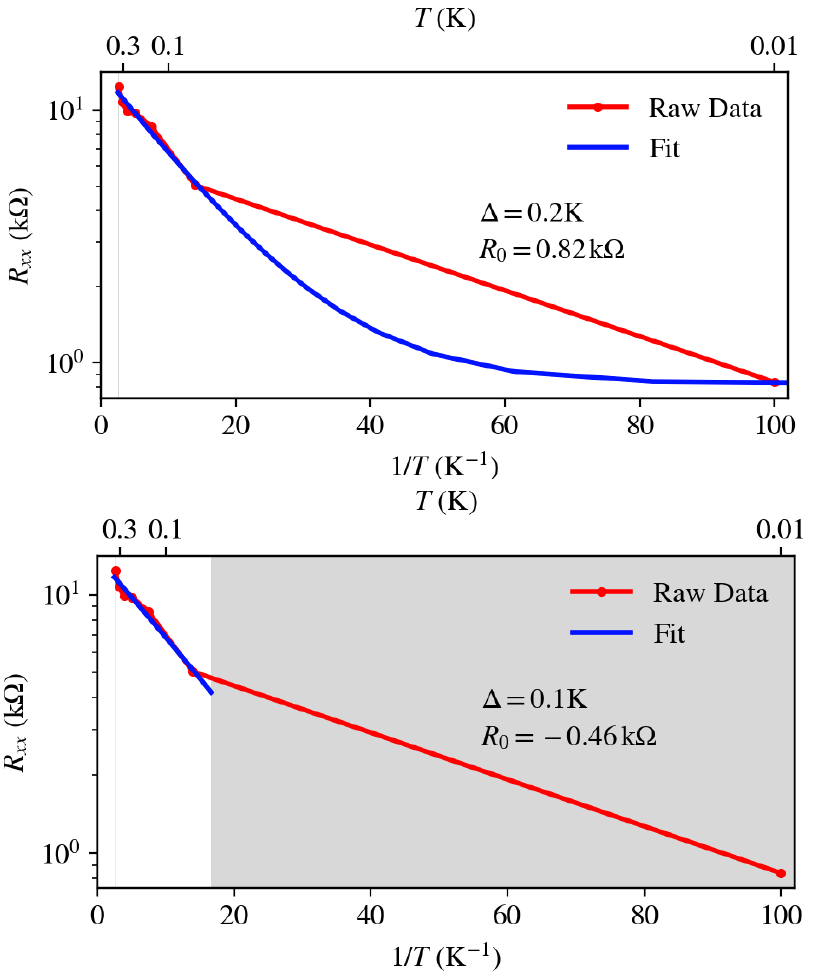}
	\caption{\label{Fit23} Thermal activation fitting of the longitudinal resistance for $\nu=2/3$. 
		The two panels correspond to fitting with different temperature ranges.
		In the lower panel, data in the shaded region are excluded from the fitting.
	}
\end{figure}

We mention that the thermal crossover 
happens not for all regions of the FQAHE states observed at higher $T$.
Interestingly, for $D/\epsilon_0$ smaller than about $0.92$ V/nm, 
the FQAHE survives to lower $T$ without any observed crossover to EQAHE.  
Also, the crossover is  dependent on the fractional states in Ref.~\onlinecite{JuEQAH2024}: 
The crossover to EQAHE occurs only for FQAHE states above half filling, \textit{i.e.}, at $\nu=3/5, 4/7, 5/9$
[as shown in Figs. 4(i)-4(k) of Ref.~\onlinecite{JuEQAH2024}],
whereas FQAHE states at $\nu=2/5$, $3/7$, $4/9$ and $5/11$ show no signature of transitioning to EQAHE. 
Although the emphasis in Ref.~\onlinecite{JuEQAH2024} (as in our work) is on the crossover from FQAHE to EQAHE  for some specific $D$ and $\nu$ values,
it is by no means reporting a generic low-$T$ absence of FQAHE.
In fact, there are large ranges of $D$ values 
where the system manifests neither QAHE nor FQAHE at any temperature, and is entirely trivial. 
Thus, the physics depends sensitively on all three parameters: $D$, $\nu$, $T$.  
The crucial dependence on $D$,
as discussed in depth later in this work,
is understandable  since each $D$ defines a unique sample 
with its own band structure and screening properties;
tuning $D$ changes the sample in a continuous manner 
with different topological properties and effective disorder strengths.
Just as in regular 2D QHE/FQHE (e.g. 2D GaAs subjected to high magnetic fields)
different samples manifest different quantum Hall properties, 
with some manifesting FQHE and others just QHE (and still others just SI), 
here changing $D$ in a single sample accomplishes the same physics, 
and the thermal crossover occurs only for some, but not for all, $D$ values.

\begin{figure}[b!]
	\includegraphics[width=0.35\textwidth]{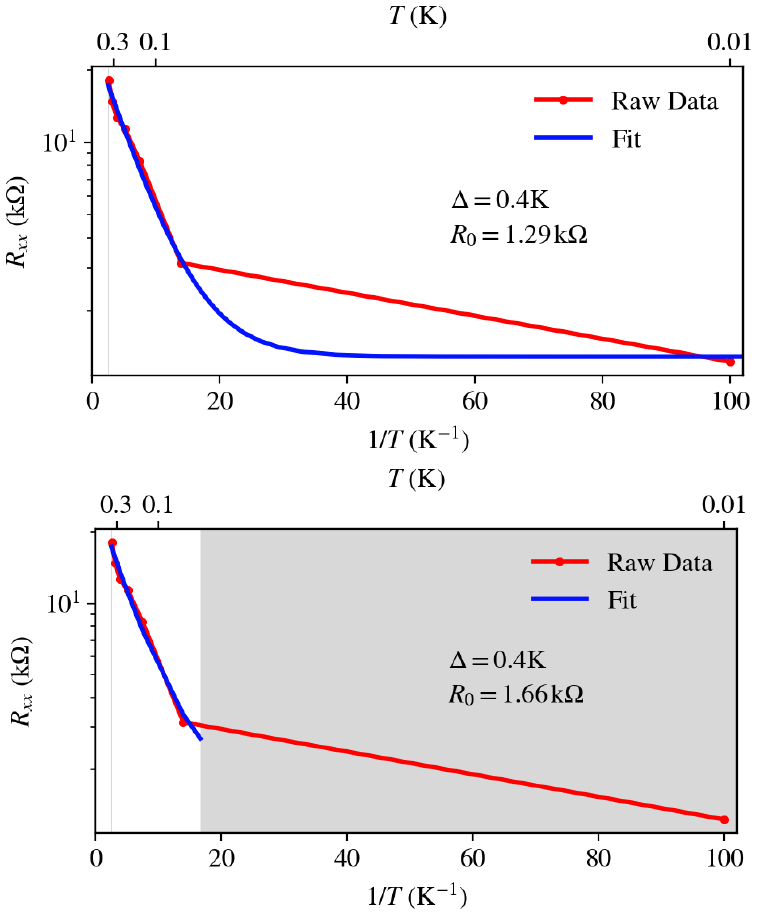}
	\caption{\label{Fit35} Thermal activation fitting of the longitudinal resistance for $\nu=3/5$. 
		The same convention is used as in Fig.~\ref{Fit23}.
	}
\end{figure}

\begin{figure}[b!]
	\includegraphics[width=0.35\textwidth]{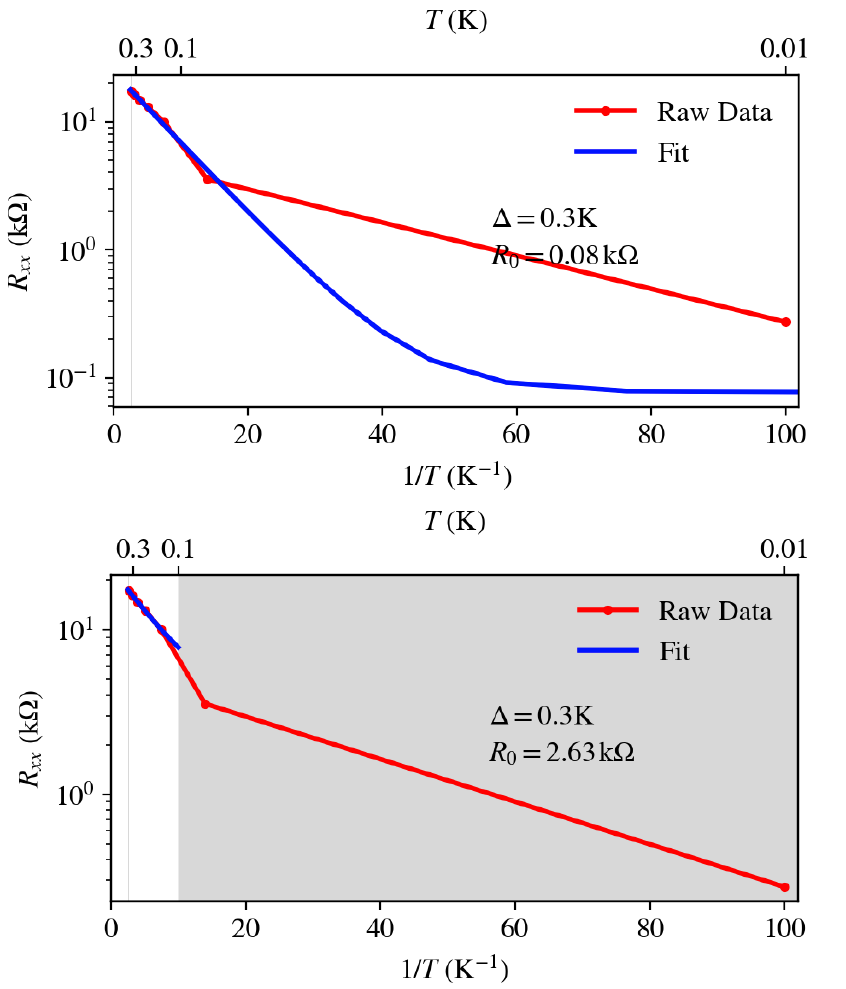}
	\caption{\label{Fit12} Thermal activation fitting of the longitudinal resistance for $\nu=1/2$.
		The same convention is used as in Fig.~\ref{Fit23}. 
	}
\end{figure}

To further emphasize the crossover, we show in Fig.~\ref{LowTRxy} the explicit $R_{xy} (T)$ extracted from Ref.~\onlinecite{JuEQAH2024} 
corresponding to the $R_{xx} (T)$ presented in Fig.~\ref{LowTRxx}.  
While the $R_{xy} (T)$ at $\nu=1$ remains quantized within $0.1\%$ of $h/e^2$,  
the results at fractional fillings emphatically bring out 
the crossover phenomenon: a low-$T$ QAHE for $\nu=2/3, 3/5$ is becoming a high-$T$ FQAHE for $T\sim 0.3$ K. 
Although at the same $D$ value, the $R_{xy} (T)$ for $\nu=1/2$ does not approach any quantized value.
(At a slightly smaller $D$, it approaches the value $2h/e^2$  expected for the composite Fermi liquid phase \cite{West1993,Read1993}.)
We note that Fig.~\ref{LowTRxy} 
indicates 
that the experimental temperature range for the fractional fillings is dominated by crossover physics 
(as we emphasized above in the context of Fig.~\ref{LowTRxx}) 
with neither the EQAHE nor the FQAHE being well established at low and high temperatures respectively 
since the quantization is hardly exact at any $T$ \cite{footnote},
except at $\nu=1$, where the system remains a well-quantized QAHE with a large activation gap and vanishing $R_{xx}$ throughout.  
This crossover behavior from low-$T$ EQAHE to high-$T$ FQAHE (in contrast to a phase transition) 
is further reinforced by our direct calculation of the activation gap for the FQAHE observed in Ref.~\onlinecite{JuEQAH2024} 
for different $\nu$ values as presented below in our Figs.~\ref{Fit23}-\ref{Fit12}.

The thermal activation fits to the data of Ref.~\onlinecite{JuEQAH2024} follow the same procedure 
as what we used in our earlier work \cite{XiePenta2024} for analyzing the reported high-$T$ ($T>0.3$ K) FQAHE of Ref.~\onlinecite{PentaGraphene2023}.  
We use the relation $R_{xx}(T) = R_0 + A \exp(-\Delta / 2k_B T)$, 
where $A$ is a constant on the order of $h/e^2$ and $\Delta$ is the thermal activation gap, 
to obtain the best fit values for $R_0$ and $\Delta$
for different fillings $\nu =1/2, 3/5, 2/3$ using the low-$T$ data of Ref.~\onlinecite{JuEQAH2024}  [shown in Figs.~\ref{Fit23}-\ref{Fit12}].  
\corev{The top and bottom panels in Figs.~\ref{Fit23}–\ref{Fit12} correspond to the fits over a full and a narrow temperature window, respectively.}
These fits work over a very narrow range of $T$, consistent with the physics 
being dominated by crossover rather than by a phase (or even two phases) over the whole range of $T \in (0.01 \text{ K}, 0.3\text{ K})$. 
The values of $\Delta$ are $\sim 0.2$ K 
for all three fractional $\nu$ values in sharp contrast to the much larger $\Delta \sim 5$ K found \cite{XiePenta2024}
for the FQAHE reported in Ref.~\onlinecite{PentaGraphene2023}.  
In addition, the extracted residual resistance ($R_0$) is much smaller in Ref.~\onlinecite{JuEQAH2024}, 
being $R_0 \sim 0.2-1 \kOhm$ in sharp contrast to $R_0 > 10$ \kOhm~found for the FQAHE in Ref.~\onlinecite{JuEQAH2024}.  
On the other hand, the high-$T$ ($>0.3$ K) results in Fig.~\ref{LowTRxy} are hardly quantized 
at the expected values of $R_{xy} = h/(\nu e^2)$, 
with the deviation from the quantization being of the order of $10-15\%$ for $\nu=2/3\text{ and }3/5$, 
which is not inconsistent with having a background  residual resistance $R_0\sim0.2-1$ \kOhm\cite{footnote}.
These results decisively show that the observed FQAHE in both Refs.~\onlinecite{JuEQAH2024,PentaGraphene2023} 
are fragile intermediate phases
\corev{(between the low-$T$ EQAHE phase and the normal 
phase at even higher $T$ where the FQAHE disappears)},
and not the ground states.  
\corev{Because of the intermediate nature of the FQAHE phase, the gap measurement is most likely inaccurate 
as the accessible dynamic range of temperature is small and does not include the low temperature range,
which instead is dominated by the crossover to the EQAHE phase.}
This also solves the puzzle of why the extracted gaps for the FQAHE at different $\nu$ were all the same ($\sim 5$ K) 
as well as why the background residual resistance was large for the FQAHE in Ref.~\onlinecite{PentaGraphene2023} as analyzed in Ref.~\onlinecite{XiePenta2024}.  
This is most likely because the gap measurement is inaccurate since a very small dynamic range of temperature is explored in the data."
We emphasize that the QAHE at $\nu=1$ is, however, a real ground state topological phase and not a crossover, 
and it has a small $R_0$ and comparable activation gaps in both Refs.~\onlinecite{JuEQAH2024,PentaGraphene2023} 
although the available low-$T$ data in Ref.~\onlinecite{JuEQAH2024} are not sufficient to precisely estimate a QAHE gap at $\nu=1$; all 
we can say is that the gap at $\nu=1$ is orders of magnitude larger than $0.1$ K ($>10$ K most likely) and the associated $R_0 <100 \Ohm$.

\begin{figure}[b!]
	\includegraphics[width=0.45\textwidth]{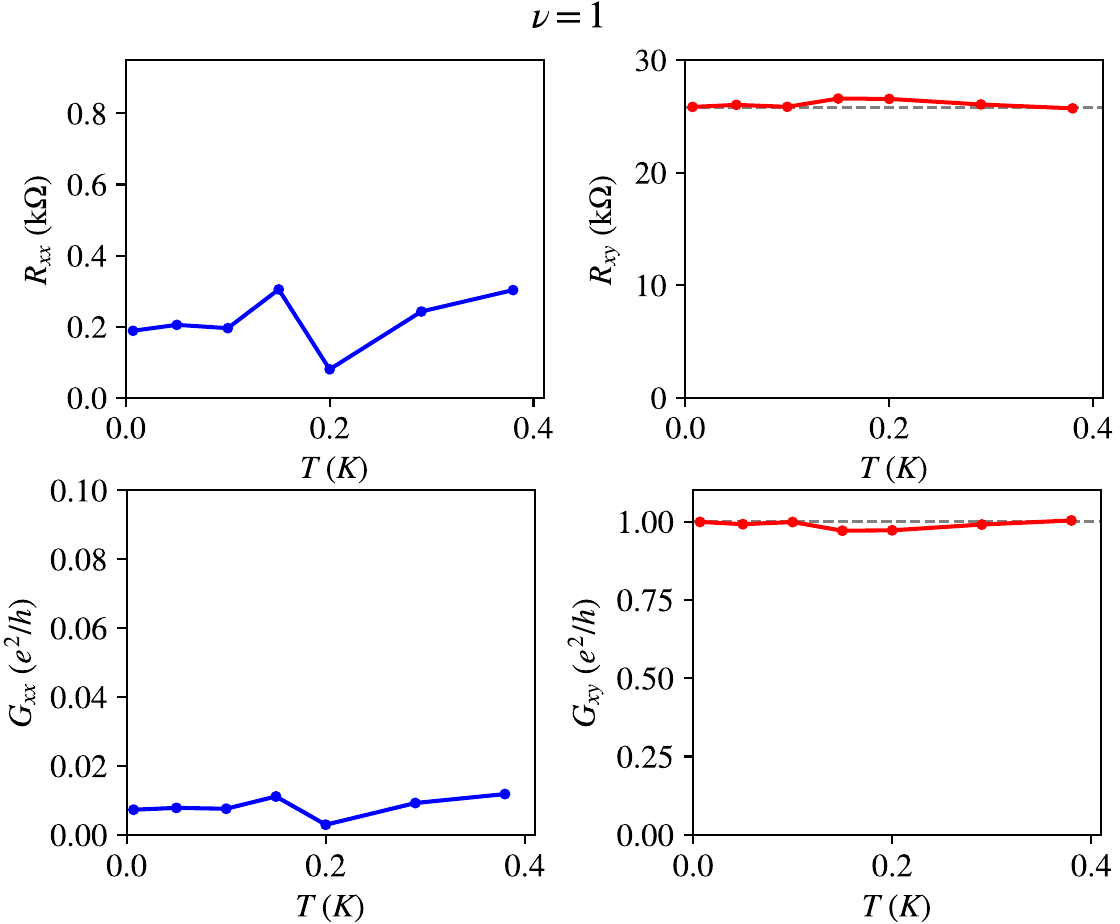}
	\caption{\label{Gxx11} Temperature dependence of the resistances and conductances at integer filling factor $\nu=1$.
		Dashed lines mark the expected quantized values.
	}
\end{figure}

\begin{figure}[t!]
	\includegraphics[width=0.45\textwidth]{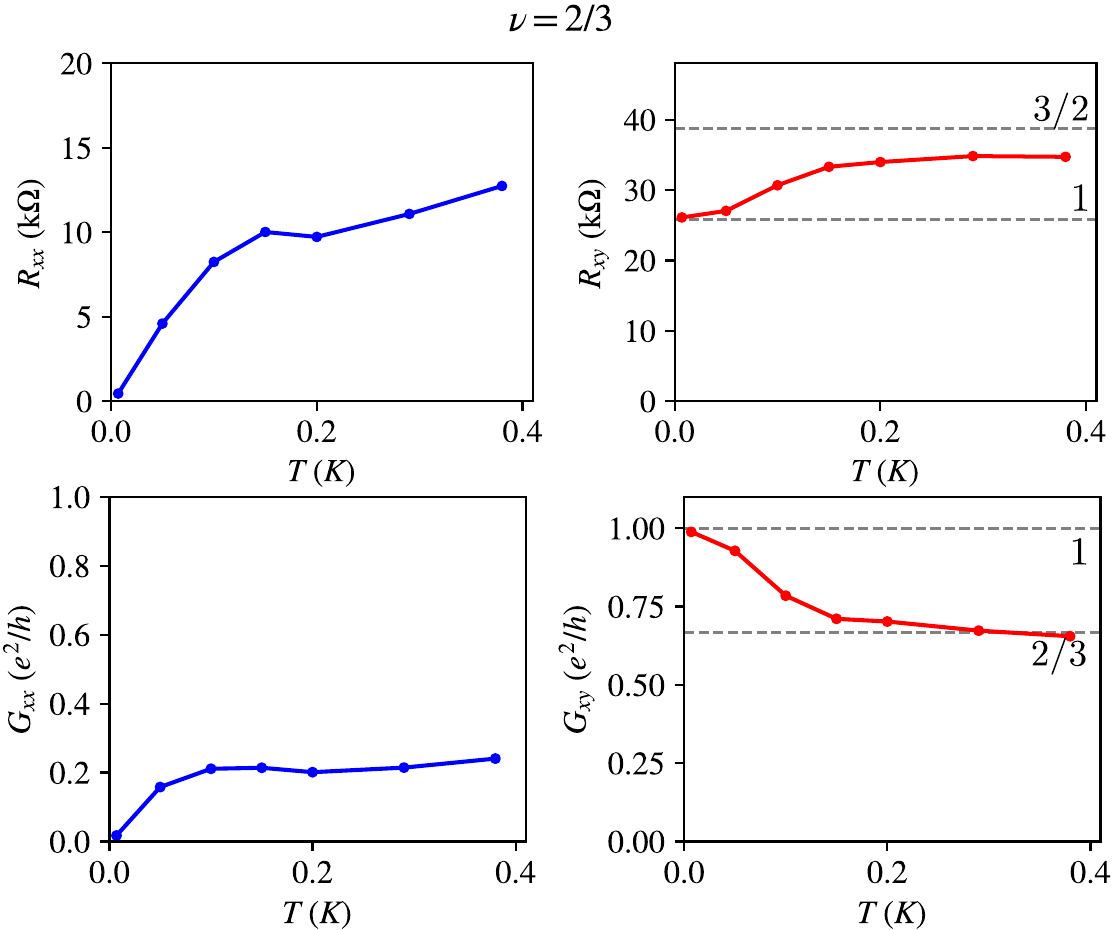}
	\caption{\label{Gxx23} Temperature dependence of the resistances and conductances at filling factor $\nu=2/3$.
		Dashed lines mark the expected quantized values.
	}
\end{figure}

\begin{figure}[b!]
	\includegraphics[width=0.45\textwidth]{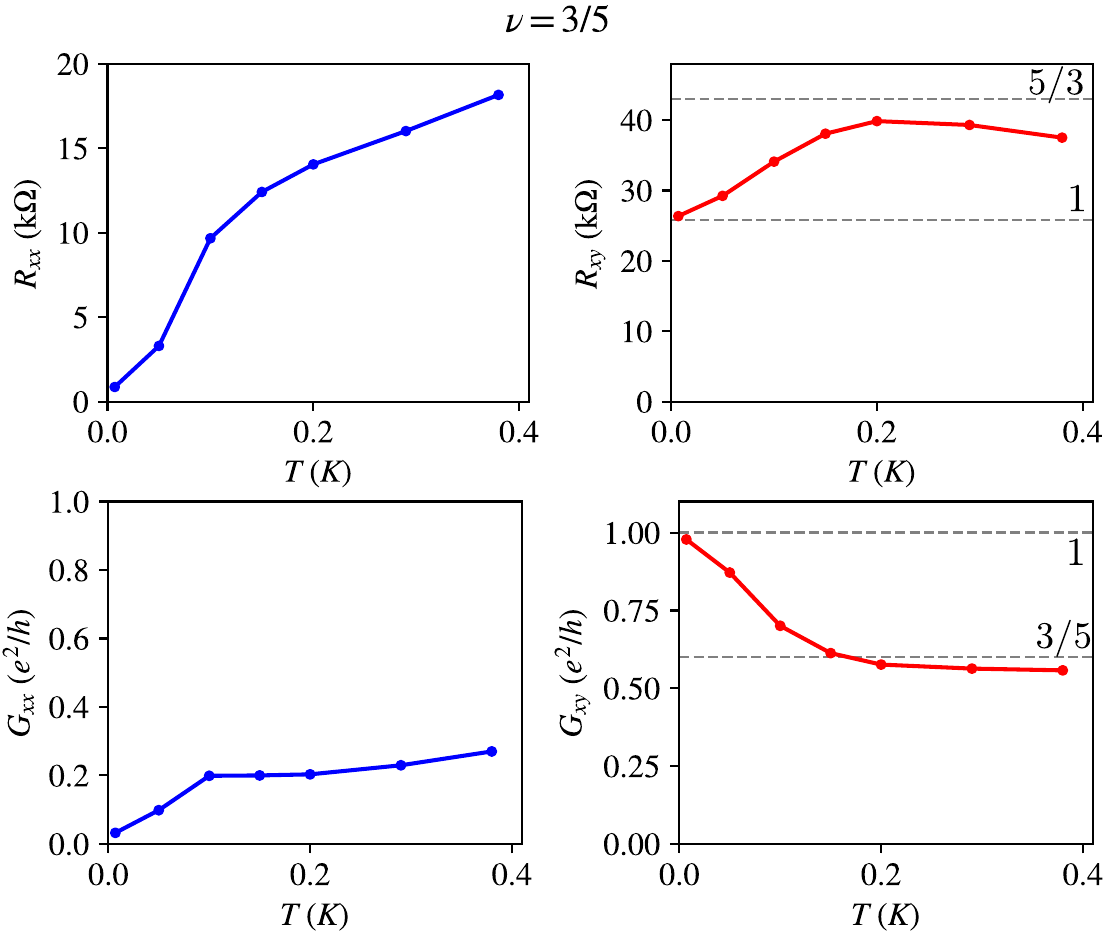}
	\caption{\label{Gxx35} Temperature dependence of the resistances and conductances at filling factor $\nu=3/5$.
		Dashed lines mark the expected quantized values.
	}
\end{figure}

\begin{figure}[t!]
	\includegraphics[width=0.45\textwidth]{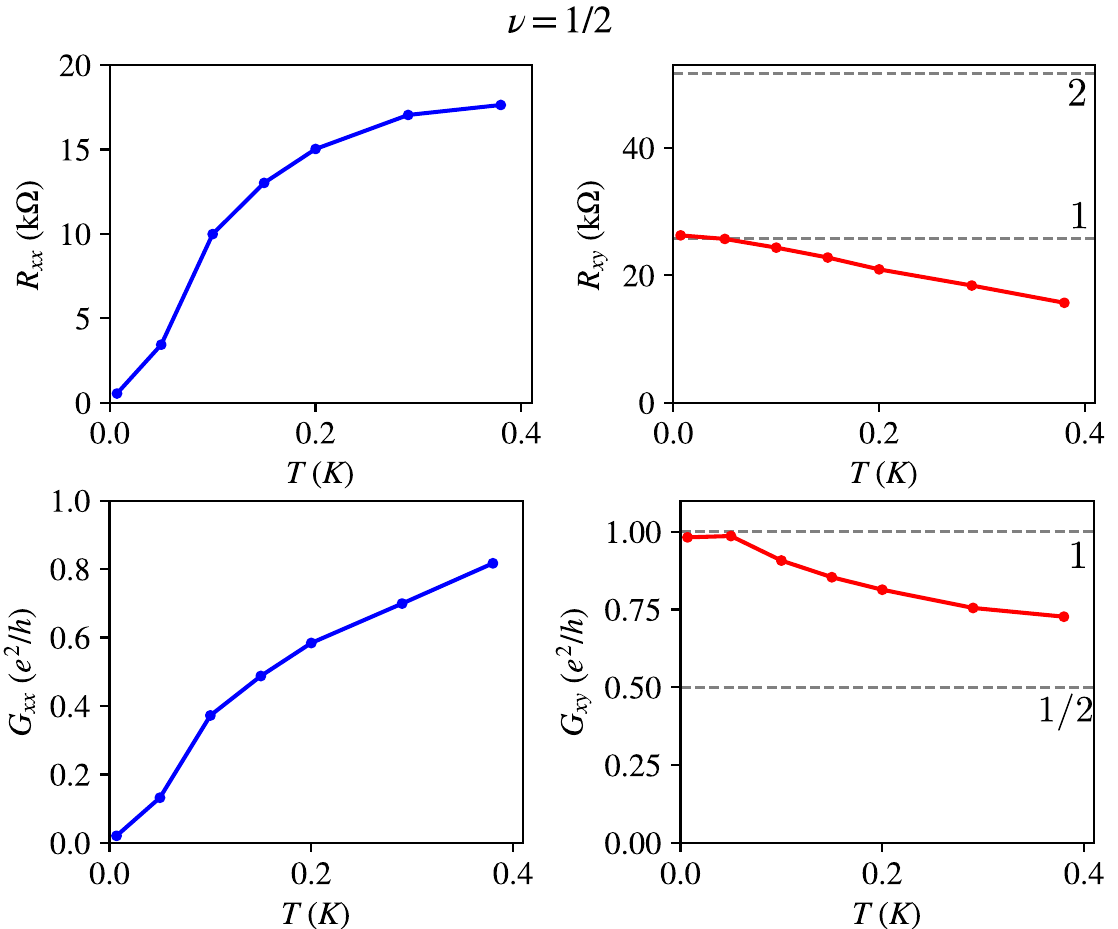}
	\caption{\label{Gxx12} Temperature dependence of the resistances and conductances at filling factor $\nu=1/2$.
		Dashed lines mark the expected quantized values.
	}
\end{figure}

We show 
in Figs.~\ref{Gxx11}-\ref{Gxx12} our calculated $T$-dependence of $R_{xx}$ and $R_{xy}$ 
as well as $G_{xx}$ and $G_{xy}$, obtained by inverting the resistance matrix $R$,
for $\nu=1, 2/3,3/5,1/2$ and at fixed $D$ \corev{where the crossover is most prominent (i.e., same value as in Fig. 2c and 2d of Ref.~\onlinecite{JuEQAH2024})}.   
These results reinforce the crossover nature of the anomalous quantum Hall phenomenology at fractional fillings, 
while also establishing that the $\nu=1$ state is indeed a stable QAHE phase.
\corev{As shown in Figs.~\ref{Gxx11}, the $\nu=1$ state
exhibits a reasonable quantization of $G_{xy}=e^2/h$ and $R_{xy} =h/e^2$ 
over the whole range of $T \in (0.01 \text{ K}, 0.3\text{ K})$ as well as having 
a rather small background resistance $R_0 \sim 100\,\Omega$ appropriate for the quantum Hall quantization.   
The fractional states in Figs.~\ref{Gxx23}-\ref{Gxx12}, on the other hand, manifest strong crossover behavior
with $G_{xy}$ varying from $e^2/h$ at very low $T$ 
to within $15\%$ of the exact value $\nu e^2/h$ at higher $T$ 
(except at $\nu=1/2$ which is not close to any quantization at this value of $D$),
 over the same temperature range. 
We emphasize that the physics is strictly crossover physics at fractional fillings, 
and only the QAHE phase at $\nu=1$ is a real phase with a true ground state quantization.  
}


\begin{figure*}[t!]
	\includegraphics[width=0.9\textwidth]{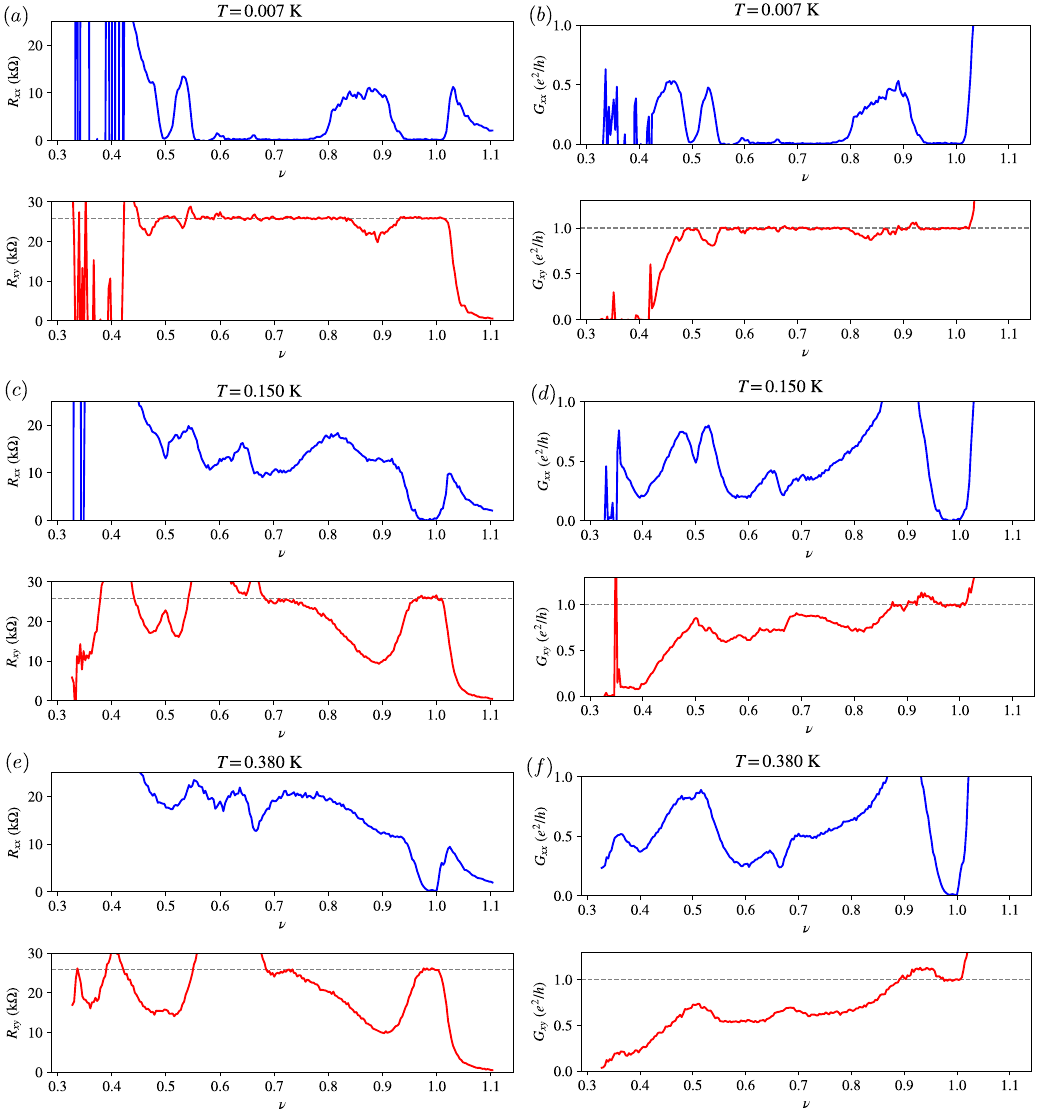}
	\caption{\label{nudependence} Filling factor dependence of the resistances and conductances at fixed temperatures 
		(a-b) $T=0.07$ K, (c-d) $T=0.15$ K and (e-f) $T=0.38$ K.
		Dashed lines mark the unit of resistance/conductance.
		Data are from Figs. 2(c)-2(d) of Ref.~\onlinecite{JuEQAH2024} (courtesy of Ju et al.).
	}
\end{figure*}

To complete the phenomenology,  
we provide in Fig.~\ref{nudependence} the low-$T$ \cite{JuEQAH2024} EQAHE to high-$T$ \cite{PentaGraphene2023} FQAHE crossover for fractional filling 
by showing our calculated $\nu$-dependence of $R_{xx}$ and $R_{xy}$ as well as $G_{xx}$ and $G_{xy}$ 
for fixed temperatures (between $7$ mK base temperature to $380$ mK) at the same $D$ value as in Figs.~\ref{Gxx11}-\ref{Gxx12}.
Note that this analyses use data exclusively from Ref.~\onlinecite{JuEQAH2024} and 
no attempt is made to reconcile the raw data between Refs.~\onlinecite{JuEQAH2024,PentaGraphene2023}, 
particularly since the temperature calibration in Ref.~\onlinecite{PentaGraphene2023} most likely has errors \cite{footnote}.
(Note also that the very low temperature measurements in these data from Ref.~\onlinecite{JuEQAH2024} 
most likely refer only to the base temperature and not necessarily the electron temperature \cite{LongPrivate}.)
What is obvious from Fig.~\ref{nudependence} is that at the lowest $T$, 
the system is almost entirely in the EQAHE phase from $\nu\sim 0.5$ to $1.0$, 
and for $\nu< 2/5$, the system is a strong insulator (where $G_{xx} \sim 0$, but $R_{xx}$ is large).  
As $T$ increases, strong crossover effects set in, but the underlying EQAHE phase survives up to $0.1$ K.  
For $T>0.1$ K, signatures for FQAHE start developing, but they are neither robust nor definitive 
when one looks at the totality of the analyzed data over the whole temperature and filling ranges.

Having established that the observed phenomenology of Ref.~\onlinecite{JuEQAH2024} is 
a low-$T$ to high-$T$ crossover between the EQAHE and the FQAHE originally reported in Ref.~\onlinecite{PentaGraphene2023}, 
we now address the possible cause underlying this crossover.  
We first mention that this crossover is surprising because the FQAHE is considered to be 
an interaction-induced fragile phase which typically manifests itself at low temperatures whereas here \cite{JuEQAH2024,PentaGraphene2023}
it is in fact disappearing at the lowest temperatures while existing at higher temperatures 
(or rather at intermediate temperatures, since FQAHE obviously disappears at very high temperatures again 
when $T>\Delta_{\nu}$).
Although this phenomenology of an FQHE existing only in the intermediate temperatures, 
while disappearing at the lowest temperatures, is unusual, it is not unheard of.  
For example, it often happens in the regular high-field fractional quantum Hall effect (FQHE), 
but only for very low density samples where the low fractions 
in the lowest Landau level (LL) are involved in the physics \cite{ManfraP,ShayeganP}. 
For example, Fig. 2 in Ref.~\onlinecite{Shayegan2022}
gives an example of a $1/7$ FQHE state arising in a 2D GaAs electron system 
(with a low carrier density of  $6\times10^{10}$ cm$^{-2}$) only in an intermediate temperature range $\sim 90$ mK 
and vanishing at lower (and higher) temperatures.  
In addition, this $1/7$ FQHE occurs on a highly resistive ($R_{xx}\sim 10$~M$\Omega$!)  background, 
which is rather similar to the FQAHE in Ref.~\onlinecite{PentaGraphene2023} happening with a large resistance, $R_0>10$ \kOhm.  
Similarly, Fig. 3(a) in Ref.~\onlinecite{Shayegan2023}
gives an example of an intermediate temperature ($T\sim 125$ mK) FQHE at $3/13$ filling 
in a low density ($\sim 10^{11}$ cm$^{-2}$) 2D GaAs hole system, 
which disappears at both lower ($T\sim 100$ mK) and higher ($T\sim 150$ mK) temperatures, 
appearing only in the intermediate temperature range similar to 
the FQAHE phenomenology in PLG being discussed in the current work.  
This $3/13$ intermediate temperature FQHE in the 2D hole system also happens in
a very large background $R_{xx} \sim 10$ M$\Omega$.  
Thus, the phenomenology of a higher-$T$ FQHE disappearing at lower temperatures 
along with a large background resistance is not unheard of in the regular FQHE literature.  

What is new in the PLG FQAHE is that the low-$T$ phase turns out to be an EQAHE phase 
for some $\nu$ values whereas at other $\nu$ values ($\nu<2/5$), 
the PLG manifests a highly resistive insulating phase similar to that in Refs.~\onlinecite{Shayegan2022,Shayegan2023}. 
In the regular high-field FQHE literature (e.g. 2D GaAs), 
such a thermal crossover from a high-$T$ FQHE-like phase to a low-$T$ quantum Hall effect (QHE) phase 
has never been reported to the best of our knowledge; there 
the crossover is from a high-$T$ apparent FQHE phase to a low-$T$ SI phase \cite{Shayegan2022,Shayegan2023}. 
We cannot rule out the possibility that at still lower temperatures 
the PLG manifests only an SI phase for all filling as it does now 
(at perhaps 40mK electron temperature) for $\nu<2/5$.  
In regular QHE/FQHE literature, three distinct situations exist generically 
at the lowest temperatures for 2D low-density systems in the lowest LL: 
(1) no QHE or FQHE is observed with the lowest LL being entirely an SI phase; 
(2) the lowest LL is entirely a QHE phase  with perhaps a transition to the SI phase at the lowest filling; 
(3) the lowest LL manifests FQHE at some fillings transitioning to the SI phase at sample dependent values of very low fillings.  
We emphasize that the temperature-induced QHE/FQHE dichotomies manifested in Refs.~\onlinecite{Shayegan2022,Shayegan2023} and in Refs.~\onlinecite{JuEQAH2024,PentaGraphene2023} 
are not ground state phases but temperature induced crossover physics as we emphasize in the current work.  
The ultimate $T=0$ phase in the PLG is likely to be an SI phase for all filling or a EQAHE phase for all filling, 
but more lower-$T$ experiments are needed to settle this issue.

What could be the reason for the observed thermal crossover?  
We speculate that the crossover arises from a competition between disorder and interaction.  
If the effective disorder is very strong (as it must be for $\nu<2/5$ in Ref.~\onlinecite{JuEQAH2024}) the ground state is an SI.  
The fact that strong disorder suppresses  FQHE \cite{Haldane1985, Haldane2003, DasSarma1985}
as well as FQAHE \cite{DasSarma2012}, 
leading to a strongly insulating phase, is theoretically known in the literature, and is of course well-established experimentally.
For lower effective disorder (as is happening for $\nu>2/5$), the ground state is QAHE 
(manifesting as EQAHE for $\nu<1$ and QAHE for $\nu \sim 1$). 
The reason the system crosses over to an FQAHE phase at slightly higher (but not very high) temperatures 
is most likely because some fraction of the localized electrons forming the $T=0$ SI phase are thermally activated into delocalized carriers, 
and as they start interacting with each other, 
they may form an intermediate FQAHE phase (perhaps as happening in Refs.~\onlinecite{Shayegan2022,Shayegan2023} in the high-field FQHE).  
Whether such a phenomenon happens or not is a delicate function of disorder, screening, interaction, and temperature.  
The phenomenology also depends sensitively on the filling factor and the displacement field 
since the screening properties of the carriers depend sensitively on these two,
and what matters is the screened effective disorder instead of the bare disorder.  
The fact that the screened disorder in the quantum Hall systems 
depends sensitively on the filling factor has been well-known for a long time \cite{SDS1981,SDS1990}. 
It is obvious that the effective disorder must depend on the applied displacement field 
since the whole PLG bandstructure depends sensitively on the applied field.  
Thus, there are two separate energy scales in the problem associated 
with the effective disorder strength ($E_{dis}$) and the effective interaction strength ($E_{int}$), 
leading to the observed crossover phenomena.  
SI and/or EQAHE dominate for $E_{int}<T<E_{dis}$, 
but there may be an optimal fine-tuned situation where $E_{dis}<T<E_{int}$ 
allowing a crossover from the low-$T$ EQAHE to the intermediate-$T$ FQAHE.  
For even higher $T$, when $E_{int}<T$ (or more precisely the temperature exceeds the FQAHE excitation gap), 
the FQAHE itself disappears.  

The existence of two distinct energy scales $E_{dis}$ and $E_{int}$, 
arising from two distinct terms in the Hamiltonian (\textit{i.e.}, disorder and interaction), 
is what may lead to the competition necessary for the crossover physics.  
The reason that this crossover is much more prominent in PLG Chern insulators 
than in regular 2D high-field QHE/FQHE physics is most likely because PLG has three different knobs 
allowing for the fine-tuning of this competition and crossover: 
filling factor $\nu$, displacement field $D$, and temperature $T$
(the temperature is an important scale because the screening is typically strongly $T$-dependent).
By contrast, regular 2D high-field QHE/FQHE has only the filling factor as the tuning knob, 
making it highly unlikely that this competition between $E_{dis}$ and $E_{int}$ 
can be fine tuned except at very low densities where disorder effects are the strongest leading to very large $R_{xx}$ values.
The importance of the displacement field in PLG is 
 directly signified by 
the $D$ dependence in the experimental observation of QAHE, FQAHE, and EQAHE in Refs.~\onlinecite{JuEQAH2024,PentaGraphene2023}.  
It is possible that if the disorder could be fine tuned at will 
(independent of the filling factor $\nu$) 
in regular 2D high-field LL systems, then the QHE to FQHE crossover 
could be observed at arbitrary filling factors in 2D systems too \cite{JainPrivate}.  
The lack of such an additional tuning knob makes this crossover possible 
only at very low $\nu$ in 2D Landau level systems, 
and thus in 2D systems the QHE to FQHE crossover as $T$ increases has not been reported 
although the crossover from SI to FQHE has been occasionally seen at very low densities\cite{ManfraP,ShayeganP,Shayegan2022,Shayegan2023}. 

We note that our conclusion for the observed EQAHE-to-FQAHE thermal crossover in PLG,
arising from a competition between interaction and disorder, is agnostic about 
whether the PLG EQAHE ground state is a Wigner crystal (the so-called Hall crystal \cite{Halperin1989}) or not, 
as has been speculated in several \cite{Dong2024,Parker2024,Sheng2024,Dong2023,Zhang2023, Senthil2023}
but not all \cite{Xiao2024,FanZhang2024,Bernevig4,Bernevig3,Bernevig2,Liu2023} recent publications.  
The Hall crystal is essentially a Wigner crystal which is pinned by disorder and for all practical purposes, 
it is simply a localized insulator for our discussion.  
For the Wigner crystal possibility, the energy scale $E_{dis}$ could also include the possible melting temperature of the Wigner crystal \cite{Vu2022}, 
which also must depend sensitively on the filling factor $\nu$ and the displacement field D as discussed above.  
We emphasize that any actual calculation of the energy scales $E_{dis}$ and $E_{int}$ 
(and therefore, of the crossover temperature scale for the EQAHE to FQAHE crossover) 
is essentially impossible at this point since the bare disorder itself is unknown in PLG 
and even calculating the energy scales for the pristine system (i.e. $E_{int}$ and $\Delta$) is a challenge 
and is highly controversial since the results depend crucially on the approximation scheme 
as emphasized in Ref.~\onlinecite{FanZhang2024} from a microscopic perspective 
and in Ref.~\onlinecite{XiePenta2024} from a phenomenological perspective.  
More experiments and considerable theoretical work are necessary 
for a better understanding of the quantitative aspects of the QAHE to FQAHE crossover observed in Ref.~\onlinecite{JuEQAH2024}, 
but the basic physics of the thermal crossover arising from a competition 
among interaction, disorder, and temperature is reasonably clear.

A very rough estimate of the crossover temperature scale from the EQAHE at lower $T$ to the FQAHE at higher $T$ 
reported in Ref.~\onlinecite{JuEQAH2024} can be obtained if we assume 
that the crossover is arising specifically from the thermal melting of an underlying Wigner crystal, 
i.e., if we accept the theoretical premise that the $T=0$ ground state of the PLG Chern insulator is an anomalous Hall crystal \cite{Halperin1989}
with $R_{xy} = h/e^2$ for all $\nu$ and $R_{xx}=0$ (due to the crystal being pinned by disorder) 
as claimed in some of the recent theoretical works\cite{Dong2024,Parker2024,Sheng2024,Dong2023}. 
We note that the anomalous Hall crystal ground state produces an integer Hall resistance quantization 
with $R_{xy}=h/e^2$ throughout the whole range of $\nu$ from 0 to 1, 
and as such, the observation of the FQAHE in Refs.~\onlinecite{JuEQAH2024,PentaGraphene2023} at higher temperatures is inconsistent 
with a Hall crystal ground state, but is consistent with our crossover scenario 
if the crystal happens to melt around the crossover temperature scale ($100-300$ mK in Figs.~\ref{Gxx11}-\ref{Gxx12} above) 
where the gradual transition from the EQAHE to FQAHE occurs.  

There is no available estimate of the thermal melting temperature of an anomalous Hall crystal in the literature, 
but there are estimates for the thermal melting temperature of the 2D Wigner crystal \cite{Vu2022,Huang2024,Hwang2001, Anderson1979, Jain2013}. 
Using the available literature for the melting of the 2D Wigner crystal 
(and assuming that such estimates apply to the 2D Hall crystal in PLG), 
we find that any existing PLG-based anomalous Hall crystal would have a melting temperature of $50-500$ mK, 
depending on the precise carrier density.  
Interestingly, this estimate is roughly consistent with the observed EQAHE-FQAHE crossover scale $\sim 0.1-0.3$ K in Ref.~\onlinecite{JuEQAH2024}.  
This could very well be a coincidence since the melting temperature of the PLG anomalous Hall crystal is theoretically unknown.  
In addition, the pristine Hall crystal must be pinned by disorder, 
and hence in the end the crossover may have nothing to do with the melting of the Hall crystal 
and everything to do with the strength of the disorder.  
Also, the observed EQAHE ground state clearly depends both on the filling fraction $\nu$ and the applied displacement field $D$, 
which argue for the importance of the effective disorder playing a key role through screening.  
Unfortunately, nothing is known about the amount of disorder in the experimental PLG samples, 
so there is no clue on how to estimate the activation gap for the disorder-induced localization 
(or equivalently, pinning of the Hall crystal)  in the sample. 

Future temperature dependent QAHE/FQAHE experiments on additional  PLG samples 
would be highly desirable in this context since the Hall crystal induced crossover phenomenology 
should be fairly universal for all samples (at similar carrier densities and displacement fields) 
whereas the disorder effects should vary randomly from sample to sample, 
thus producing different crossover behaviors in different samples.  
It may be useful here to mention the corresponding situation in the regular 2D high-field LL QHE/FQHE experiments, 
where the physics depends crucially on disorder, 
and there is no universal behavior for samples manifesting QHE or FQHE;
everything depends on the sample quality (characterized by the sample mobility), 
and in general, higher mobility samples manifest both QHE and FQHE, 
and lower mobility samples manifest only QHE, and very poor quality samples do not manifest even QHE in the lowest LL.  
A classic example is the original discovery \cite{Klitzing1980} of QHE by von Klitzing in the 2D Si-SiO$_2$ MOSFETs, 
where the QHE manifested only in the higher LL and the lowest spin- and valley-split LL 
only manifested strong localization with $R_{xx}$ ($G_{xx}$) being very large (small).  
In fact, FQHE has never been observed in Si MOSFETs because of their high disorder level.  
Only the highest mobility 2D GaAs (and graphene) samples typically 
show many FQHE fractions in the lowest LL at high magnetic field, 
with the system eventually crossing over from the FQHE ground state to a SI ground state at very low filling fractions ($\sim1/7$), 
which is sometimes interpreted as a transition to the Wigner crystal phase, 
but could very well be simply a strongly localized Anderson insulator 
arising from the effective disorder becoming very strong at low carrier densities \cite{Tsui1988,Cunningham1990}.
It is interesting to note in this context that Refs.~\onlinecite{JuEQAH2024,PentaGraphene2023} also manifest a strongly localized 
insulating phase (for $\nu<2/5$) over a large regime of $D$ and $\nu$ values, 
indicating the importance of disorder in PLG QAHE/FQAHE physics.

Finally, while we have focused on the intriguing temperature-induced FQAHE to QAHE 
crossover phenomenon reported in Ref.~\onlinecite{JuEQAH2024}, 
we mention the interesting low-temperature, displacement-field-dependent 
feature observed in the same work (see Figs.~4(i)-4(k) therein).
The displacement field induces a quantum phase transition between FQAHE and QAHE 
for several fractional $\nu$s at the lowest temperature ($T\sim 10$ mK).  
Such a direct quantum phase transition between FQHE and QHE has never been reported 
in the high-field 2D quantum Hall literature (e.g. GaAs), 
but is theoretically allowed here because the $D$ field modifies the band structure 
(and thus, the quantum metric), enabling the phase transition as shown in Figs.~4(i)-4(k) of Ref.~\onlinecite{JuEQAH2024}.
This is thus akin to something like changing the effective mass (or perhaps the effective interaction) 
in the corresponding regular high-field QHE problem
(which is not experimentally feasible since there is no available knob for continuously tuning the effective mass or interaction).
On the other hand, changing $D$ also modifies the effective disorder, 
which can induce a phase transition between the SI phase and the FQAHE/EQAHE phase, 
as also shown in Figs.~4(i)-4(k) and Figs.~4(a)-4(b) of Ref.~\onlinecite{JuEQAH2024} at lower $D$ values.  
Such SI-FQHE transitions are often reported in regular high-field 2D QHE systems (as we have discussed above already) by varying $\nu$, 
and are interpreted as Wigner crystallization or Anderson localization.  
Very interestingly, in the zero-field PLG system, 
these transitions between WC/AI phases and QAHE/FQAHE phases can be induced 
by tuning either $\nu$ or $D$.  
The tunability by the displacement field allows for a much richer phenomenology in PLG
than in the high-field 2D systems.
We emphasize that although there is no temperature induced quantum phase transition in Ref.~\onlinecite{JuEQAH2024}, 
$D$ and $\nu$ induced quantum phase transitions at $T=0$ are indeed observed in Refs. \onlinecite{JuEQAH2024} and \onlinecite{PentaGraphene2023}.
We also note  that  the observed FQAHE to QAHE thermal crossover is limited to the $\nu>1/2$ regime 
whereas the $\nu<1/2$ regime, manifesting weak FQAHE for some $\nu$ values (e.g., $\nu=2/5, 3/7, 4/9, 5/11$), 
is increasingly dominated by the strongly localized SI states.
This leads to the interesting question of whether the ground state may involve an SI phase
for $\nu<1/2$ and a QAHE phase for $\nu>1/2$ 
with a possible $T=0$ quantum phase transition from SI to QAHE at $\nu=1/2$, 
which is theoretically plausible. 
Only future low-temperature experiments can address this crucial question.

\emph{Note added.} The following preprints [\onlinecite{XiaoLi2024},\onlinecite{Senthil2024}] appeared after
the submission of our work.

\begin{acknowledgments}
	{\em Acknowledgment.}---\noindent	This work is supported by the Laboratory for Physical Sciences through the Condensed Matter Theory Center at the University of Maryland.  The authors thank Long Ju and Zhengghuang Lu for providing the experimental transport data needed for the theoretical analyses, and also for numerous helpful discussions on the details of the experiments in Refs.~\onlinecite{JuEQAH2024,PentaGraphene2023}.  The authors thank Mike Manfra and Mansour Shayegan for helpful correspondence on fractional quantum Hall effects in low-density GaAs systems. The authors thank Jainendra Jain and Andrei Bernevig for helpful discussions. One of the authors (S.D.S) acknowledges the hospitality of the Aspen Center for Physics (partially funded by the National Science Foundation) where the basic idea underlying this work was formulated during a  2024 summer program.
\end{acknowledgments}


\begin{thebibliography}{99}
	
\bibitem{JuEQAH2024}
Z. Lu, T. Han, Y. Yao, J. Yang, J. Seo, L. Shi, S. Ye, K. Watanabe, T. Taniguchi, and L. Ju, 
Extended quantum anomalous Hall states in graphene/hBN moir\'e superlattices, 
arXiv:2408.10203.

\bibitem{PentaGraphene2023}
Z. Lu, T. Han, Y. Yao, A. P. Reddy, J. Yang, J. Seo, K. Watanabe, T. Taniguchi, L. Fu, and L. Ju,
Fractional quantum anomalous Hall effect in a graphene moir\'e superlattice,
Nature (London) \textbf{626}, 759 (2024).


\bibitem{XiePenta2024}
M. Xie and S. Das Sarma,
Integer and fractional quantum anomalous Hall effects in pentalayer graphene, 
Phys. Rev. B \textbf{109}, L241115 (2024).

\bibitem{Jain1989}
J. K. Jain, 
Composite-fermion approach for the fractional quantum Hall effect, 
Phys. Rev. Lett. \textbf{63}, 199 (1989).


\bibitem{footnote}
While the $R_{xy}$ is not fully quantized at $T\sim 0.3$ K, it is so in the previous report \cite{PentaGraphene2023}.
This discrepancy is likely due to the difference in the calibration of electronic temperature
as improvement in electronic temperature control is made in the follow-up work \cite{JuEQAH2024, LongPrivate}. 


\bibitem{West1993}
R. R. Du, H. L. Stormer, D. C. Tsui, L. N. Pfeiffer, and K. W. West, 
Experimental evidence for new particles in the fractional quantum Hall effect, 
Phys. Rev. Lett. \textbf{70}, 2944 (1993).

\bibitem{Read1993}
B. I. Halperin, P. A. Lee, and N. Read, 
Theory of the half-filled Landau level, 
Phys. Rev. B \textbf{47}, 7312 (1993).

\bibitem{LongPrivate}
L. Ju, (private communication).

\bibitem{ManfraP}
M. Manfra, (private communication).

\bibitem{ShayeganP}
M. Shayegan, (private communication).

\bibitem{Shayegan2022}
Y. J. Chung, D. Graf, L. W. Engel, K. A. Villegas Rosales, P. T. Madathil, K. W. Baldwin, K. W. West, L. N. Pfeiffer, and M. Shayegan, 
Correlated states of 2D electrons near the Landau level filling $\nu=1/7$, 
Phys. Rev. Lett. \textbf{128}, 026802 (2022)

\bibitem{Shayegan2023}
C. Wang, A. Gupta, S. K. Singh, P. T. Madathil, Y. J. Chung, L. N. Pfeiffer, K. W. Baldwin, R. Winkler, and M. Shayegan, 
Fractional quantum Hall state at filling factor $\nu=1/4$ in ultra-high-quality GaAs two-dimensional hole systems, 
Phys. Rev. Lett. \textbf{131}, 266502 (2023).

\bibitem{Haldane2003}
D. N. Sheng, X. Wan, E. H. Rezayi, K. Yang, R. N. Bhatt, and F. D. M. Haldane, 
Disorder-driven collapse of the mobility gap and transition to an insulator in the fractional quantum Hall effect, 
Phys. Rev. Lett. \textbf{90}, 256802 (2003)

\bibitem{Haldane1985}
E. H. Rezayi and F. D. M. Haldane, 
Incompressible states of the fractionally quantized Hall effect in the presence of impurities: A finite-size study, 
Phys. Rev. B \textbf{32}, 6924(R) (1985).

\bibitem{DasSarma1985}
F. C. Zhang, V. Z. Vulovic, Y. Guo, and S. Das Sarma, 
Effect of a charged impurity on the fractional quantum Hall effect: Exact numerical treatment of finite systems, 
Phys. Rev. B \textbf{32}, 6920(R) (1985).

\bibitem{DasSarma2012}
S. Yang, K. Sun, and S. Das Sarma, 
Quantum phases of disordered flatband lattice fractional quantum Hall systems, 
Phys. Rev. B \textbf{85}, 205124 (2012).


\bibitem{SDS1981}
S. Das Sarma, 
Two-dimensional level broadening in the extreme quantum limit, 
Phys. Rev. B \textbf{23}, 4592 (1981).

\bibitem{SDS1990}
X. C. Xie, Q. P. Li, and S. Das Sarma, 
Density of states and thermodynamic properties of a two-dimensional electron gas in a strong external magnetic field, 
Phys. Rev. B \textbf{42}, 7132 (1990).

\bibitem{JainPrivate}
J. Jain, (private communication).

\bibitem{Halperin1989}
Z. Tešanović, F. Axel, and B. I. Halperin, 
``Hall crystal" versus Wigner crystal, 
Phys. Rev. B \textbf{39}, 8525 (1989).

\bibitem{Dong2024}
Z. Dong, A. S. Patri, and T. Senthil, 
Stability of Anomalous Hall Crystals in multilayer rhombohedral graphene, 
arXiv: 2403.07873.

\bibitem{Parker2024}
T. Soejima, J. Dong, T. Wang, T. Wang, M. P. Zaletel, A. Vishwanath, and D. E. Parker, 
Anomalous Hall Crystals in Rhombohedral Multilayer Graphene II: General Mechanism and a Minimal Model, 
arXiv:2403.05522.

\bibitem{Sheng2024}
D. N. Sheng, A. P. Reddy, A. Abouelkomsan, E. J. Bergholtz, and L. Fu, 
Quantum anomalous Hall crystal at fractional filling of moiré superlattices, 
arXiv:2402.17832.

\bibitem{Dong2023}
J. Dong, T. Wang, T. Wang, T. Soejima, M. P. Zaletel, A. Vishwanath, and D. E. Parker, 
Anomalous Hall Crystals in Rhombohedral Multilayer Graphene I: Interaction-Driven Chern Bands and Fractional Quantum Hall States at Zero Magnetic Field, 
arXiv:2311.05568.

\bibitem{Zhang2023}
B. Zhou, H. Yang, and Y.-H. Zhang, 
Fractional quantum anomalous Hall effects in rhombohedral multilayer graphene in the moiréless limit and in Coulomb imprinted superlattice, 
arXiv:2311.04217.

\bibitem{Senthil2023}
Z. Dong, A. S. Patri, and T. Senthil, 
Theory of fractional quantum anomalous Hall phases in pentalayer rhombohedral graphene moiré structures, 
arXiv:2311.03445. 

\bibitem{Xiao2024}
K. Huang, S. Das Sarma, and X. Li, 
Fractional quantum anomalous Hall effect in rhombohedral multilayer graphene with a strong displacement field, 
arXiv:2408.05139.

\bibitem{FanZhang2024}
K. Huang, X. Li, S. Das Sarma, and F. Zhang, 
Self-consistent theory for the fractional quantum anomalous Hall effect in rhombohedral pentalayer graphene, 
arXiv:2407.08661.

\bibitem{Bernevig4}
J. Yu, J. Herzog-Arbeitman, Y. H. Kwan, N. Regnault, and B. A. Bernevig, 
Moir\'e Fractional Chern Insulators IV: Fluctuation-Driven Collapse of FCIs in Multi-Band Exact Diagonalization Calculations on Rhombohedral Graphene, 
arXiv:2407.13770.

\bibitem{Bernevig3}
Y. H. Kwan, J. Yu, J. Herzog-Arbeitman, D. K. Efetov, N. Regnault, and B. A. Bernevig, 
Moir\'e Fractional Chern Insulators III: Hartree-Fock Phase Diagram, Magic Angle Regime for Chern Insulator States, the Role of the Moiré Potential and Goldstone Gaps in Rhombohedral Graphene Superlattices, 
arXiv:2312.11617.

\bibitem{Bernevig2}
J. Herzog-Arbeitman, Y. Wang, J. Liu, P. M. Tam, Z. Qi, Y. Jia, D. K. Efetov, O. Vafek, N. Regnault, H. Weng, Q. Wu, B. A. Bernevig, and J. Yu, 
Moir\'e fractional Chern insulators. II. First-principles calculations and continuum models of rhombohedral graphene superlattices, 
Phys. Rev. B \textbf{109}, 205122 (2024).

\bibitem{Liu2023}
Z. Guo, X. Lu, B. Xie, and J. Liu, 
Theory of fractional Chern insulator states in pentalayer graphene moiré superlattice, 
arXiv:2311.14368.

\bibitem{Vu2022}
D. Vu and S. Das Sarma, 
Thermal melting of a quantum electron solid in the presence of strong disorder: Anderson localization versus the Wigner crystal,
Phys. Rev. B \textbf{106}, L121103 (2022).

\bibitem{Huang2024}
Y. Huang and S. Das Sarma, 
Electronic transport, metal-insulator transition, and Wigner crystallization in transition metal dichalcogenide monolayers, 
Phys. Rev. B \textbf{109}, 245431 (2024).

\bibitem{Hwang2001}
E. H. Hwang and S. Das Sarma, 
Plasmon dispersion in dilute two-dimensional electron systems: Quantum-classical and Wigner crystal–electron liquid crossover, 
Phys. Rev. B \textbf{64}, 165409 (2001).

\bibitem{Anderson1979}
H. Fukuyama, P. M. Platzman, and P. W. Anderson, 
Two-dimensional electron gas in a strong magnetic field, 
Phys. Rev. B \textbf{19}, 5211 (1979).

\bibitem{Jain2013}
A. C. Archer, K. Park, and J. K. Jain, 
Competing crystal phases in the lowest Landau level, 
Phys. Rev. Lett. 111, 146804 (2013).


\bibitem{Klitzing1980}
K. v. Klitzing, G. Dorda, and M. Pepper, 
New method for high-accuracy determination of the fine-structure constant based on quantized Hall resistance, 
Phys. Rev. Lett. \textbf{45}, 494 (1980).

\bibitem{Tsui1988}
V. J. Goldman, M. Shayegan, and D. C. Tsui, 
Evidence for the fractional quantum Hall state at $\nu=1/7$, 
Phys. Rev. Lett. \textbf{61}, 881 (1988).

\bibitem{Cunningham1990}
V. J. Goldman, M. Santos, M. Shayegan, and J. E. Cunningham, 
Evidence for two-dimensional quantum Wigner crystal, 
Phys. Rev. Lett. \textbf{65}, 2189 (1990).

\bibitem{XiaoLi2024}
K. Huang, S. Das Sarma, and X. Li, Impurity-induced thermal crossover in fractional Chern insulators, arXiv:2409.04349.

\bibitem{Senthil2024}
A. S. Patri, Z. Dong, and T. Senthil, Extended quantum anomalous Hall effect in moiré structures: phase transitions and transport, arXiv:2408.11818.


\end{thebibliography}
\end{document}